\documentstyle[12pt]{article}

\begin{document}

\topmargin 0pt
\oddsidemargin 5mm

\setcounter{page}{0}
\begin{titlepage}
\vspace{2cm}
\begin{center}
{{\large\bf
 PERTURBATION THEORY IN ANGULAR QUANTIZATION APPROACH AND THE EXPECTATION
VALUES OF EXPONENTIAL FIELDS IN SIN-GORDON MODEL
}}\\

\vspace{1cm}

{\large R.H.Poghossian}\\
\vspace{1cm}

{\em Yerevan Physics Institute,  Republic of Armenia}\\
{(Alikhanian Brothers St. 2,  Yerevan 375036,  Armenia)}\\
{e-mail: poghos@moon.yerphi.am}
\end{center}

\vspace{5mm}

\centerline{{\bf{Abstract}}}

In angular quantization approach a perturbation theory for the Massive
Thirring Model (MTM) is developed, which allows us to calculate Vacuum
Expectation Values of exponential fields in sin-Gordon theory near the free
fermion point in first order of MTM coupling constant $g$. The
Hankel-transforms play an important role when carrying out this
calculations. The expression we have found coincides with that of the 
direct expansion over $g$ of the exact formula conjectured by S.Lukyanov and
A.Zamolodchikov.
 
\vfill
\centerline{\large Yerevan Physics Institute}
\centerline{\large Yerevan 1999}

\vfill
\end{titlepage}

\vspace{0.5cm} \setcounter{equation}{0}

\section{Introduction}

\renewcommand{\theequation}{1.\arabic{equation}}

Sin-Gordon theory is one of the most studied examples of exactly integrable
relativistic QFT in two dimensions. Its action 
\begin{equation}
{\cal S}_{SG}=\int d^{2}x\left\{ \frac{1}{16\pi }\partial _{\nu }\varphi
\partial ^{\nu }\varphi +2\mu \cos \beta \varphi \right\}  \label{0.5}
\end{equation}
may be viewed as a perturbation of the free massless Bose field by the relevant
operator $\cos \beta \varphi $ with scaling dimension $d=2\beta ^{2}<2.$ It
is convenient to normalize exponential fields by the condition, that in UV
limit their two point functions approach to those of $c=1$ conformal free
Bose field theory 
\begin{equation}
\left\langle e^{ia\varphi \left( x\right) }e^{-ia\varphi \left( y\right)
}\right\rangle _{SG}\rightarrow \left| x-y\right| ^{-4a^{2}}\quad as\ \left|
x-y\right| \rightarrow 0.  \label{0.6}
\end{equation}
The on-shell properties of this theory i.e. the mass spectrum and the
S-matrix are well known \cite{ZamZam1}. The lightest particles of the theory
are solitons and antisolitons in term of which sin-Gordon theory has an
equivalent Lagrangian formulation with the action of Massive Thirring Model
(MTM) \cite{Col} 
\begin{equation}
{\cal S}_{MTM}=\int d^{2}x\left\{ i\overline{\Psi }\gamma ^{\nu }\partial
_{\nu }\Psi -M\ \overline{\Psi }\Psi -\frac{g}{2}\left( \overline{\Psi }
\gamma ^{\nu }\Psi \right) \left( \overline{\Psi }\gamma _{\nu }\Psi \right)
\right\} ,  \label{0.7}
\end{equation}
where $\overline{\Psi }$, $\Psi $ are two component Dirac spinors and the
corresponding (anti-) particles are identified with the (anti-) solitons of ( 
\ref{0.5}). The famous Coleman relations 
\begin{equation}
\frac{g}{\pi }=\frac{1}{2\beta ^{2}}-1;\ J^{\nu }\equiv \overline{\Psi }
\gamma ^{\nu }\Psi =-\frac{\beta }{2\pi }\epsilon ^{\nu \mu }\partial _{\mu
}\varphi  \label{0.8}
\end{equation}
serve as a dictionary between bosonic and fermionic languages. More recently
comparing the results of the thermodynamic Bethe-Ansatz analyses with those
of the Conformal Perturbation Theory an exact relation between the soliton
mass $M$ and the perturbation parameter $\mu $ is established \cite{AlZam} 
\begin{equation}
\mu =\frac{\Gamma \left( \beta ^{2}\right) }{\pi \Gamma \left( 1-\beta
^{2}\right) }\left[ \frac{M\sqrt{\pi }\Gamma \left( \frac{1+\xi }{2}\right) 
}{2\Gamma \left( \frac{\xi }{2}\right) }\right] ^{2-2\beta ^{2}},
\label{0.9}
\end{equation}
where 
\begin{equation}
\xi =\frac{\beta ^{2}}{1-\beta ^{2}}.  \label{0.10}
\end{equation}

Starting from the expressions for the special cases $\beta \rightarrow 0$
(semiclassical limit) and $\beta ^{2}=1/2$ (free fermion case) S.Lukyanov
and A.Zamolodchikov in \cite{LukZam} the following general formula for the
Vacuum Expectation Value (VEV) $G_{a}=\left\langle \exp ia\varphi \left(
0\right) \right\rangle $ have conjectured 
\begin{eqnarray}
&&G_{a}=\left( \frac{m\Gamma \left( \frac{1+\xi }{2}\right) \Gamma \left( 
1-\frac{\xi }{2}\right) }{4\sqrt{\pi }}\right) ^{2a^{2}}\times  \label{0.1}
\\
&&\exp \left\{ \int\limits_{0}^{\infty }\frac{dt}{t}\left[ \frac{\sinh
^{2}\left( 2a\beta t\right) }{2\sinh \beta ^{2}t\sinh t\cosh \left(
\left( 1-\beta ^{2}\right) t\right) }-2a^{2}e^{-2t}\right] \right\} . 
\nonumber
\end{eqnarray}
In the subsequent papers \cite{FatLukZamZam1}, \cite{FatLukZamZam2} some
convincing arguments have been presented, showing that the VEV's $\widetilde{
G}_{a}=\left\langle \exp a\varphi \left( 0\right) \right\rangle _{sh-G}$ in
sinh-Gordon theory, the action of which formally can be obtained simply
replacing $\beta \rightarrow ib$ in (\ref{0.5}), obey the functional
relations 
\begin{equation}
\widetilde{G}_{a}=R\left( a\right) \widetilde{G}_{Q-a}=\widetilde{G}_{-a},
\label{0.2}
\end{equation}
where $Q=b+1/b$ and $R\left( a\right) $ is related to the Liouville
reflection amplitude $S\left( p\right) $ \cite{ZamZam2} 
\begin{equation}
R\left( \frac{Q}{2}+ip\right) =S\left( p\right) =-\left( \frac{\pi \mu
\Gamma \left( b^{2}\right) }{\Gamma \left( 1-b^{2}\right) }\right) ^{-\frac{
2ip}{b}}\frac{\Gamma \left( 1+2ip/b\right) \Gamma \left( 1+2ipb\right) }{
\Gamma \left( 1-2ip/b\right) \Gamma \left( 1-2ipb\right) }\ .  \label{0.3}
\end{equation}
It is natural to expect that $\widetilde{G}_{a}$ can be obtained from $G_{a}$
making the analytic continuation 
\begin{equation}
\beta \rightarrow ib;\ a\rightarrow -ia.  \label{0.4}
\end{equation}
And indeed it can be shown that after substitution (\ref{0.4}), the
expression (\ref{0.1}) obeys the relation (\ref{0.2}). Unfortunately the
functional relations (\ref{0.2}) alone are not sufficient to find $
\widetilde{G}_{a}$ uniquely: the multiplication by any even, periodic with
period $Q$ function of $a$ gives a different solution to (\ref{0.2}).
However, $\widetilde{G}_{a}$ obtained from (\ref{0.1}) is the only {\it 
meromorphic }solution, obeying the extra requirement of "minimality" (i.e. the
condition that only the poles and zeros, imposed by the functional relations
(\ref{0.2}) are allowed). So, any independent test of the
Lukyanov-Zamolodchikov formula (\ref{0.1}) will support the minimality
assumption as well. This is important also because there are other
interesting models for which some functional relations like (\ref{0.2})
are hold and the minimality condition makes it possible to find exact VEV's.

In this article a perturbation theory based on the angular quantization \cite
{Luk} of the MTM (\ref{0.7}) is developed, using which
Lukyanov-Zamolodchikov formula (\ref{0.1}) near the free fermion point in
first order over the MTM coupling constant $g$ is tested.

The section 2 is devoted to the angular quantization of the MTM (\ref{0.7}).
A particular attention is payed to the local field product regularization
procedure, which has some additional features in comparison with the case of
the ordinary quantization in Cartesian coordinates.

In section 3 we calculate VEV $\left\langle \exp ia\varphi \left( 0\right)
\right\rangle $ near the free fermion point. It appears that the
Hankel-transform is a very useful tool to carry out this calculation. Some
of the related mathematical details are presented in Appendix A. The choices
we have made to regularize the traces over the fermionic Fock space and the
local field product, result in a finite multiplicative renormalization of
the field $\exp ia\varphi \left( 0\right) $. The corresponding
renormalization factor is calculated using the methods of Boundary CFT \cite
{Cardy} in Appendix B. The final expression we obtained for the expansion of
the VEV $\left\langle \exp ia\varphi \left( 0\right) \right\rangle $ up to
the first order in MTM coupling constant $g$ is in complete agreement with
the Lukyanov-Zamolodchikov conjecture (\ref{0.1}).

\vspace{0.5cm} \setcounter{equation}{0}

\section{Angular Quantization of the Massive Thirring Model}

\renewcommand{\theequation}{2.\arabic{equation}}

It is convenient to use the ciral representation of the Dirac matrices

\begin{equation}
\gamma ^{0}=\sigma _{2}=\left( 
\begin{array}{ll}
0 & -i \\ 
i & 0
\end{array}
\right) ,\ \gamma ^{1}=-i\sigma _{1}=\left( 
\begin{array}{ll}
0 & -i \\ 
-i & 0
\end{array}
\right)  \label{1.1}
\end{equation}
and denote the components of Dirac spinors as

\begin{equation}
\Psi \equiv \left( 
\begin{array}{l}
\psi _{L} \\ 
\psi _{R}
\end{array}
\right) ,\ \overline{\Psi }\equiv \Psi ^{\dagger }\gamma ^{0}.  \label{1.2}
\end{equation}
In this notations the action (\ref{0.7}) in Euclidean space reads

\begin{eqnarray}
{\cal A}_{MTM} &=&\int d^{2}z\ 2\left[ \psi _{R}^{\dagger }\partial \psi
_{R}+\psi _{L}^{\dagger }\overline{\partial }\psi _{L}-\frac{iM}{2}\left(
\psi _{L}^{\dagger }\psi _{R}-\psi _{R}^{\dagger }\psi _{L}\right) \right. 
\nonumber \\
&&\ \qquad \qquad \qquad \qquad \qquad \qquad \qquad \quad \left. +g\psi
_{L}^{\dagger }\psi _{L}\psi _{R}^{\dagger }\psi _{R}\right] ,  \label{1.3}
\end{eqnarray}
where $z=\tau +ix^{1}$, $\overline{z}=\tau -ix^{1}$ are complex coordinates
on the plane ($\tau =ix^{0}$ is the Euclidean time ), $\partial \equiv
\partial /\partial z$, $\overline{\partial }\equiv \partial /\partial 
\overline{z}$ and the measure $d^{2}z\equiv dx^{1}d\tau $.

As the VEV's of local fields $\left\langle e^{ia\varphi \left( 0\right)
}\right\rangle $ have rotational symmetry, it is natural to use the
conformal polar coordinates $\eta $, $\theta $ defined by

\begin{equation}
z\equiv e^{\eta +i\theta };\quad \overline{z}\equiv e^{\eta -i\theta }
\label{1.4}
\end{equation}
and treat $\eta $, $\theta $ as space and (Euclidean) time respectively.
Under the coordinate transformation (\ref{1.4}) the Fermi fields transform
as follows:

\begin{equation}
\psi _{L}\left( z,\overline{z}\right) = e^{-\frac{i\pi }{4}-\frac{ \eta
+i\theta }{2}}\psi _{L}\left( \eta ,\theta \right) ;\ \psi _{R}\left( z, 
\overline{z}\right) = e^{\frac{i\pi }{4}-\frac{\eta -i\theta }{2} }\psi
_{R}\left( \eta ,\theta \right) ,  \label{1.5}
\end{equation}
and similarly for the conjugate fields $\psi _{L}^{\dagger }$, $\psi
_{R}^{\dagger }$. In this coordinates the action (\ref{1.3}) takes the form:

\begin{eqnarray}
{\cal A}_{MTM} &=&\int\limits_{0}^{2\pi }d\theta \int\limits_{-\infty
}^{\infty }d\eta \ \left[ \psi _{L}^{\dagger }\left( \partial _{\theta
}-i\partial _{\eta }\right) \psi _{L}+\psi _{R}^{\dagger } \left( \partial
_{\theta}+i\partial _{\eta }\right) \psi _{R}-\right.  \nonumber \\
\quad \qquad \qquad &&\qquad \quad \left. iMe^{\eta }\left( \psi
_{L}^{\dagger }\psi _{R}-\psi _{R}^{\dagger }\psi _{L}\right) +2g\psi
_{L}^{\dagger }\psi _{L}\psi _{R}^{\dagger }\psi _{R}\right] .  \label{1.6}
\end{eqnarray}
The usual canonical quantization yields the following ''equal time''
anti-commutation relations:

\begin{equation}
\left\{ \psi _{L}\left( \eta \right) ,\psi _{L}^{\dagger }\left( \eta
^{\prime }\right) \right\} =\delta \left( \eta -\eta ^{\prime }\right)
,\quad \left\{ \psi _{R}\left( \eta \right) ,\psi _{R}^{\dagger }\left( \eta
^{\prime }\right) \right\} =\delta \left( \eta -\eta ^{\prime }\right) .
\label{1.7}
\end{equation}
From (\ref{1.6}) one deduces that the Hamiltonian defining the evolution
along $\theta $ is given by 
\begin{eqnarray}
\mbox{{\bf K}} &=&\int\limits_{-\infty }^{\infty }d\eta \left[ -\psi
_{L}^{\dagger }i\partial _{\eta }\psi _{L}+\psi _{R}^{\dagger }i\partial
_{\eta }\psi _{R}-iMe^{\eta }\left( \psi _{L}^{\dagger }\psi _{R}-\psi
_{R}^{\dagger }\psi _{L}\right) +\right.  \nonumber \\
&&\qquad \qquad \qquad \qquad \qquad \qquad \qquad \quad \left. 2g\psi
_{L}^{\dagger }\psi _{L}\psi _{R}^{\dagger }\psi _{R}\right] .  \label{1.8}
\end{eqnarray}
To regularize the theory, following \cite{LukZam} let us restrict the range
of the ''space'' coordinate $\eta $ to the semi-infinite box $\eta \in
\left[ \log \varepsilon ;\infty \right) $ with $M\varepsilon \ll 1$ and
impose the boundary conditions 
\begin{equation}
\psi _{L}+\psi _{L}^{\dagger }\mid _{\eta =\log \varepsilon }=\psi _{R}+\psi
_{R}^{\dagger }\mid _{\eta =\log \varepsilon }=0.  \label{1.9}
\end{equation}
As usual, to develope perturbation theory in interaction picture, one first
has to diagonalize the quadratic part of the Hamiltonian (\ref{1.8}). This
can be achieved using the decomposition \cite{LukZam} 
\begin{eqnarray}
&&\psi _{L}\left( \eta ,\theta \right) =\sum\limits_{\upsilon \in {\cal N}
_{\varepsilon }}\frac{1}{\sqrt{2\pi \rho \left( \nu \right) }}\ c_{\nu
}u_{\nu }\left( \eta \right) e^{-\nu \theta }\ ,  \nonumber \\
\ &&\psi _{R}\left( \eta ,\theta \right) =\sum\limits_{\upsilon \in {\cal N}
_{\varepsilon }}\frac{1}{\sqrt{2\pi \rho \left( \nu \right) }}\ c_{\nu
}\upsilon _{\nu }\left( \eta \right) e^{-\nu \theta }\ ,  \nonumber \\
&&\psi _{L}^{\dagger }\left( \eta ,\theta \right) =\sum\limits_{\upsilon \in 
{\cal N}_{\varepsilon }}\frac{1}{\sqrt{2\pi \rho \left( \nu \right) }}\
c_{\nu }^{\dagger }u_{\nu }^{*}\left( \eta \right) e^{\nu \theta }\ , 
\nonumber \\
&&\ \psi _{R}^{\dagger }\left( \eta ,\theta \right) =\sum\limits_{\upsilon
\in {\cal N}_{\varepsilon }}\frac{1}{\sqrt{2\pi \rho \left( \nu \right) }}\
c_{\nu }^{\dagger }\upsilon _{\nu }^{*}\left( \eta \right) e^{\nu \theta }\ ,
\label{1.10}
\end{eqnarray}
where the set of admissable $\nu$'s ${\cal N}_{\varepsilon }$ and the density 
of states $\rho \left( \nu \right) $ are specified below (see (\ref{1.14}), 
(\ref{1.14'})) and the wave functions \cite{LukZam} 
\begin{equation}
\left( 
\begin{array}{l}
u_{\nu }\left( \eta \right) \\ 
v_{\nu }\left( \eta \right)
\end{array}
\right) =\frac{\sqrt{2M}e^{\frac{\eta }{2}}}{\Gamma \left( \frac{1}{2}-i\nu
\right) }\left( \frac{M}{2}\right) ^{-i\nu }\left( 
\begin{array}{l}
K_{\frac{1}{2}-i\nu }\left( Me^{\eta }\right) \\ 
K_{\frac{1}{2}+i\nu }\left( Me^{\eta }\right)
\end{array}
\right) ,  \label{1.11}
\end{equation}
are solutions of the free Dirac equation ($K_{\nu }\left( x\right) $ is the
MacDonald function). The wave functions (\ref{1.11}) have the asymptotic
behavior 
\begin{equation}
\left( 
\begin{array}{l}
u_{\nu }\left( \eta \right) \\ 
v_{\nu }\left( \eta \right)
\end{array}
\right) \rightarrow \left( 
\begin{array}{l}
1 \\ 
0
\end{array}
\right) e^{i\nu \eta }+S_{F}\left( \nu \right) \left( 
\begin{array}{l}
0 \\ 
1
\end{array}
\right) e^{-i\nu \eta }\ \ as\ \eta \rightarrow -\infty \ ,  \label{1.12}
\end{equation}
where 
\begin{equation}
S_{F}\left( \nu \right) =\left( \frac{M}{2}\right) ^{-2i\nu }\frac{\Gamma
\left( \frac{1}{2}+i\nu \right) }{\Gamma \left( \frac{1}{2}-i\nu \right) }
\label{1.13}
\end{equation}
is the fermion scattering amplitude off the ''mass barrier'' \cite{LukZam}.
It follows from (\ref{1.9}), (\ref{1.12}) that the set ${\cal N}
_{\varepsilon }$ of admissable $\nu $'s over which the sum in (\ref{1.10})
is carried out consists of the solutions of the equation 
\begin{equation}
2\pi \left( n+\frac{1}{2}\right) =2\nu \log \frac{1}{\varepsilon }+\frac{1}{
i }\log S_{F}\left( \nu \right) ,  \label{1.14}
\end{equation}
where $n$ is arbitrary integer. Therefore the density of states is given by 
\begin{equation}
\rho \left( \nu \right) =\frac{dn}{d\nu }=\frac{1}{\pi }\log \frac{1}{
\varepsilon }+\frac{1}{2\pi i}\log ^{^{\prime }}S_{F}\left( \nu \right) ,
\label{1.14'}
\end{equation}
Below we'll use the notation ${\cal N}_{\varepsilon }^{+}\subset {\cal N}
_{\varepsilon }$ for the subset of positive $\nu $'s. The operators $c_{\nu
} $, $c_{\nu }^{\dagger }$ satisfy the anti-commutation relations 
\begin{eqnarray}
\left\{ c_{\nu },c_{\nu ^{\prime }}\right\} &=&\left\{ c_{\nu }^{\dagger
},c_{\nu ^{\prime }}^{\dagger }\right\} =0,  \nonumber \\
\left\{ c_{\nu },c_{\nu ^{\prime }}^{\dagger }\right\} &=&\delta _{\nu ,\nu
^{\prime }}.  \label{1.15}
\end{eqnarray}
In terms of this operators the free part of the Hamiltonian (\ref{1.8})
takes a very simple form 
\begin{equation}
{\bf K}_{0}=\sum\limits_{\nu \in {\cal N}_{\varepsilon }^{+}}\nu \left(
c_{\nu }^{\dagger }c_{\nu }+c_{-\nu }c_{-\nu }^{\dagger }\right) ,
\label{1.16}
\end{equation}
which shows that $c_{\nu }^{\dagger }$, $c_{-\nu }$ ($c_{\nu },$ $c_{-\nu
}^{\dagger }$) are fermion and anti-fermion creation (annihilation)
operators. Therefore the vacuum state $\left| 0\right\rangle $ can be
defined by 
\begin{equation}
c_{\nu }\left| 0\right\rangle =c_{-\nu }^{\dagger }\left| 0\right\rangle
=0,\quad \nu \in {\cal N}_{\varepsilon }^{+},  \label{1.17}
\end{equation}
and the Hilbert space of states ${\cal H}$ is spanned over the base vectors 
\begin{equation}
\prod\limits_{\nu \in {\cal N}_{\varepsilon }^{+}}c_{\nu }^{\dagger n_{\nu
}}c_{-\nu }^{\overline{n}_{\nu }}\left| 0\right\rangle ,  \label{1.18}
\end{equation}
where $n_{\nu }\in \left\{ 0,1\right\} $ ($\overline{n}_{\nu }\in \left\{
0,1\right\} $) are the occupation numbers of fermions (anti-fermions) with
energy $\nu $.

Standard arguments show, that the expectation value of any quantity $
\left\langle {\bf X}\right\rangle $ in interacting theory (\ref{1.8}) may be
evaluated using the formula 
\begin{eqnarray}
\left\langle {\bf X}\right\rangle \equiv &&\frac{\int D\Psi D\overline{\Psi }
e^{-{\cal A}_{MTM}}{\bf X}}{\int D\Psi D\overline{\Psi }e^{-{\cal A}_{MTM}}}=
\nonumber \\
&&\frac{{\it Tr}_{{\cal H}}\left[ e^{-2\pi {\bf K}_{0}}T\left( e^{-\int {\bf 
K}_{I}\left( \eta ,\theta \right) d\eta d\theta }{\bf X}\right) \right] }{
{\it Tr}_{{\cal H}}\left[ e^{-2\pi {\bf K}_{0}}T\left( e^{-\int {\bf K}
_{I}\left( \eta ,\theta \right) d\eta d\theta }\right) \right] }\ ,
\label{1.19}
\end{eqnarray}
where ${\bf X}$ in second line of (\ref{1.19}) as well as the interaction
term of the Hamiltonian 
\begin{equation}
{\bf K}_{I}=2g \int\limits_{\log \varepsilon }^{\infty }\ N\left( \psi
_{L}^{\dagger }\psi _{L}\psi _{R}^{\dagger }\psi _{R}\right) d\eta
\label{1.20}
\end{equation}
is taken in interaction picture and $T$ indicates the time ordering
operation. In (\ref{1.20}) we introduced the notation $N\left( \cdots
\right) $ for the suitably regularized product of the local operators in
coinciding points (see below). Appearance of trace instead of conventional
vacuum matrix element in (\ref{1.19}) is due to compactification of the
''time'' $\theta $.

Now let us turn to the regularization procedure of the products of local
operators at the coinciding points. In the ordinary case of non-compactified
time one simply implies normal ordering prescription which is well known to
be equivalent to the suppression of all contractions among the fields inside
normal ordering symbol. In contrary to the vacuum matrix element, the trace
of normal ordered product of creation and annihilation operators doesn't
vanish, therefore the analogues regularization in the case of compactified
time is slightly changed. It is easy to see that in this case the correctly
regularized perturbing operator $N\left( \psi _{L}^{\dagger }\psi _{L}\psi
_{R}^{\dagger }\psi _{R}\right) $ is given by 
\begin{eqnarray}
&&N\left( \psi _{L}^{\dagger }\psi _{L}\psi _{R}^{\dagger }\psi _{R}\right)
=\psi _{L}^{\dagger }\psi _{L}\psi _{R}^{\dagger }\psi _{R}-\left\langle
\psi _{L}^{\dagger }\psi _{L}\right\rangle _{0}\psi _{R}^{\dagger }\psi
_{R}-\left\langle \psi _{R}^{\dagger }\psi _{R}\right\rangle _{0}\psi
_{L}^{\dagger }\psi _{L}+  \nonumber \\
&&\qquad \qquad \qquad \qquad \quad \left\langle \psi _{L}^{\dagger }\psi
_{R}\right\rangle _{0}\psi _{R}^{\dagger }\psi _{L}+\left\langle \psi
_{R}^{\dagger }\psi _{L}\right\rangle _{0}\psi _{L}^{\dagger }\psi _{R}+ 
\nonumber \\
&&\qquad \qquad \qquad \qquad \quad \left\langle \psi _{L}^{\dagger }\psi
_{L}\right\rangle _{0}\left\langle \psi _{R}^{\dagger }\psi
_{R}\right\rangle _{0}-\left\langle \psi _{L}^{\dagger }\psi
_{R}\right\rangle _{0}\left\langle \psi _{R}^{\dagger }\psi
_{L}\right\rangle _{0}=  \nonumber \\
&&:\psi _{L}^{\dagger }\psi _{L}\psi _{R}^{\dagger }\psi _{R}:+\left\langle
:\psi _{L}^{\dagger }\psi _{R}:\right\rangle _{0}:\psi _{R}^{\dagger }\psi
_{L}:+\left\langle :\psi _{R}^{\dagger }\psi _{L}:\right\rangle _{0}:\psi
_{L}^{\dagger }\psi _{R}:-  \nonumber \\
&&\qquad \qquad \qquad \qquad \qquad \qquad \qquad \qquad \qquad
\left\langle :\psi _{L}^{\dagger }\psi _{R}:\right\rangle _{0}\left\langle
:\psi _{R}^{\dagger }\psi _{L}:\right\rangle _{0},  \label{1.21}
\end{eqnarray}
where 
\begin{equation}
\left\langle {\bf X}\right\rangle _{0}=\frac{{\it Tr}_{{\cal H}}e^{-2\pi 
{\bf K}_{0}}{\bf X}}{{\it Tr}_{{\cal H}}e^{-2\pi {\bf K}_{0}}}\ ,
\label{1.22}
\end{equation}
for any operator ${\bf X}$, $::$ denotes the ordinary normal ordering with
respect to the mode decomposition (\ref{1.10}) and in the second equality of
(\ref{1.21} ) we have taken into account that 
\begin{equation}
\left\langle :\psi _{L}\psi _{R}:\right\rangle _{0}=\left\langle :\psi
_{L}^{\dagger }\psi _{R}^{\dagger }:\right\rangle _{0}=\left\langle :\psi
_{L}^{\dagger }\psi _{L}:\right\rangle _{0}=\left\langle :\psi _{R}^{\dagger
}\psi _{R}:\right\rangle _{0}=0.  \label{1.24}
\end{equation}
Note that, while to give a proper mining to the first expression for the
perturbing operator $N\left( \psi _{L}^{\dagger }\psi _{L}\psi _{R}^{\dagger
}\psi _{R}\right) $ (see (\ref{1.21}) ) one has to apply point splitting,
the separate terms of the second expression are already finite.

\vspace{0.5cm} \setcounter{equation}{0}

\section{VEV's of Exponential Fields}

\renewcommand{\theequation}{3.\arabic{equation}}

The Dirac fields $\Psi \left( z,\overline{z}\right) $, $\overline{\Psi }
\left( z,\overline{z}\right) $ have non-trivial monodromy with respect to
the exponential fields $\exp ia\varphi \left( 0\right) $ 
\begin{eqnarray}
\Psi \left( z,\overline{z}\right) *e^{ia\varphi \left( 0\right) } &=&e^{i
\frac{2\pi a}{\beta }}\Psi \left( z,\overline{z}\right) e^{ia\varphi \left(
0\right) }\ ,  \nonumber \\
\overline{\Psi }\left( z,\overline{z}\right) *e^{ia\varphi \left( 0\right) }
&=&e^{-i\frac{2\pi a}{\beta }}\overline{\Psi }\left( z,\overline{z}\right)
e^{ia\varphi \left( 0\right) }\ ,  \label{2.1}
\end{eqnarray}
where $*$ denotes the analytic continuation around the point $0$. The VEV $
\left\langle \exp ia\varphi \left( 0\right) \right\rangle $ can be expressed
in terms of Grassmanian functional integral \cite{LukZam} 
\begin{equation}
I\left( a\right) =\frac{\int\limits_{{\cal F}_{a}}\left[ {\cal D}\Psi {\cal D
}\overline{\Psi }\right] e^{-{\cal A}_{MTM}}}{\int\limits_{{\cal F}
_{0}}\left[ {\cal D}\Psi {\cal D}\overline{\Psi }\right] e^{-{\cal A}_{MTM}}}
,  \label{2.2}
\end{equation}
where the functional integration in the numerator is carried out over the
space ${\cal F}_{a}$ of twisted field configurations with monodromy (\ref
{2.1}). In angular quantization picture the insertion of the operator $\exp
ia\varphi \left( 0\right) $ changes the boundary conditions along ''time''
direction so that the Hilbert space remains untouched but the Hamiltonian
due to (\ref{2.1}) acquires an additional term $ -ia{\bf Q}/\beta $,
where 
\begin{equation}
{\bf Q}=\sum\limits_{\nu \in {\cal N}_{\varepsilon }^{+}}\left( c_{\nu
}^{\dagger }c_{\nu }-c_{-\nu }c_{-\nu }^{\dagger }\right)  \label{2.3}
\end{equation}
is the fermion charge operator. Thus, due to (\ref{1.19}), for the
regularized version of the functional integral (\ref{2.2}) we have 
\begin{equation}
I_{\varepsilon }\left( a,g\right) =\frac{{\it Tr}_{{\cal H}}\left[ e^{-2\pi 
{\bf K}_{0}+\frac{2\pi ia}{\beta }{\bf Q}}T\left( e^{-\int {\bf K}_{I}\left(
\eta ,\theta \right) d\eta d\theta }\right) \right] }{{\it Tr}_{{\cal H}
}\left[ e^{-2\pi {\bf K}_{0}}T\left( e^{-\int {\bf K}_{I}\left( \eta ,\theta
\right) d\eta d\theta }\right) \right] }{\it \ }.  \label{2.4}
\end{equation}
The VEV $\left\langle \exp ia\varphi \left( 0\right) \right\rangle $ can be
expressed in terms of $I_{\varepsilon }\left( a,g\right) $ as 
\begin{equation}
\left\langle e^{ia\varphi \left( 0\right) }\right\rangle
=\lim\limits_{\varepsilon \rightarrow 0}Z^{-1}\varepsilon
^{-2a^{2}}I_{\varepsilon }\left( a,g\right) ,  \label{2.5}
\end{equation}
where $Z$ is some renormalization factor. This point, as well as the
appearance of the factor $\varepsilon ^{-2a^{2}}$, which has purely CFT
origin, will be discussed later on. 

The main goal of this paper is the
evaluation of (\ref{2.4}) and (\ref{2.5}) perturbatively up to the linear
over $g$ terms 
\begin{equation}
I_{\varepsilon }\left( a,g\right) =I_{\varepsilon }\left( a,0\right) \left(
1+gI_{\varepsilon }^{1}\left( a\right) +{\cal O}\left( g^{2}\right) \right) .
\label{2.6}
\end{equation}
The calculation of 
\begin{equation}
I_{\varepsilon }\left( a,0\right) =\frac{{\it Tr}_{{\cal H}}\left[ e^{-2\pi 
{\bf K}_{0}+2\pi ia\sqrt{2}{\bf Q}}\right] }{{\it Tr}_{{\cal H}}\left[
e^{-2\pi {\bf K}_{0}}\right] }  \label{2.7}
\end{equation}
is carried out in \cite{LukZam} and the result is 
\begin{eqnarray}
&&I_{\varepsilon }\left( a,0\right) =\varepsilon ^{2a^{2}}\left\langle
e^{ia\varphi \left( 0\right) }\right\rangle \mid _{g=0}=  \nonumber \\
&&\left( \frac{M\varepsilon }{2}\right) ^{2a^{2}}\exp \left\{
\int\limits_{0}^{\infty }\frac{dt}{t}\left[ \frac{\sinh ^{2}\left( \sqrt{2}
at\right) }{\sinh ^{2} t}-2a^{2}e^{-2t}\right] \right\} ,  \label{2.22}
\end{eqnarray}
so that we'll concentrate our attention on the second term 
\begin{eqnarray}
I_{\varepsilon }^{1}\left( a\right) &=&\frac{a}{2\pi }\partial _{a}\log
I_{\varepsilon }\left( a,0\right) -  \nonumber \\
&&\frac{4\pi {\it Tr}_{{\cal H}}\left[ e^{-2\pi {\bf K}_{0}+2\pi ia\sqrt{2}
{\bf Q}}\int\limits_{\log \varepsilon }^{\infty }d\eta N\left( \psi
_{L}^{\dagger }\psi _{L}\psi _{R}^{\dagger }\psi _{R}\right) \right] }{{\it 
Tr}_{{\cal H}}\left[ e^{-2\pi {\bf K}_{0}+2\pi ia\sqrt{2}{\bf Q}}\right] }.
\label{2.8}
\end{eqnarray}
Using mode decomposition (\ref{1.10}) and evaluating traces over ${\cal H}$
in the bases (\ref{1.18}) for (\ref{2.8}) we obtain 
\begin{eqnarray}
&&I_{\varepsilon }^{1}\left( a\right) =\frac{a}{2\pi }\partial _{a}\log
I_{\varepsilon }\left( a,0\right) +\sum\limits_{\nu _{1},\nu _{2}\in {\cal N}
_{\varepsilon }^{+}}\left\{ \frac{\cosh \pi \nu _{1}\cosh \pi \nu _{2}}{\pi
^{4}\rho \left( \nu _{1}\right) \rho \left( \nu _{2}\right) }\times \right. 
\nonumber \\
&&\int\limits_{M\varepsilon }^{\infty }\left[ A_{\nu _{1}}A_{\nu _{2}}\left(
\left| K_{\frac{1}{2}+i\nu _{1}}\left( x\right) \right| ^{2}\left| K_{\frac{1
}{2}+i\nu _{2}}\left( x\right) \right| ^{2}-K_{\frac{1}{2}+i\nu
_{1}}^{2}\left( x\right) K_{\frac{1}{2}-i\nu _{2}}^{2}\left( x\right)
\right) +\right.  \nonumber \\
&&\qquad A_{\nu _{1}}A_{\nu _{2}}^{*}\left( K_{\frac{1}{2}+i\nu
_{1}}^{2}\left( x\right) K_{\frac{1}{2}+i\nu _{2}}^{2}\left( x\right)
-\left| K_{\frac{1}{2}+i\nu _{1}}\left( x\right) \right| ^{2}\left| K_{\frac{
1}{2}+i\nu _{2}}\left( x\right) \right| ^{2}\right) -  \nonumber \\
&&\qquad \qquad \qquad \qquad A_{\nu _{1}}^{0}\left( A_{\nu _{2}}-\frac{1}{2}
A_{\nu _{2}}^{0}\right) \left( K_{\frac{1}{2}+i\nu _{1}}^{2}\left( x\right)
-K_{\frac{1}{2}-i\nu _{1}}^{2}\left( x\right) \right) \times  \nonumber \\
&&\qquad \qquad \qquad \qquad \left. \left. \left( K_{\frac{1}{2}+i\nu
_{2}}^{2}\left( x\right) -K_{\frac{1}{2}-i\nu _{2}}^{2}\left( x\right)
\right) +{\bf c.c.}\right] xdx\right\} \ ,  \label{2.9}
\end{eqnarray}
where 
\begin{equation}
A_{\nu }=\frac{e^{-2\pi \nu +2i\sqrt{2}\pi a}}{1+e^{-2\pi \nu +2i\sqrt{2}\pi
a}},  \label{2.10}
\end{equation}
and $A_{\nu }^{0}\equiv A_{\nu }\mid _{a=0}$. 

The following formulae for the
integrals over $x$ included in (\ref{2.9}) are proved in Appendix A (below
and later on we'll omit vanishing in the limit $\varepsilon \rightarrow 0$
terms)
\begin{eqnarray}
&&\frac{\cosh \pi \nu _{1}\cosh \pi \nu _{2}}{\pi ^{2}}\int\limits_{M
\varepsilon }^{\infty }\left| K_{\frac{1}{2}+i\nu _{1}}\left( x\right)
\right| ^{2}\left| K_{\frac{1}{2}+i\nu _{2}}\left( x\right) \right| ^{2}xdx=
\nonumber \\
&&\int\limits_{0}^{\infty }\left[ \frac{\sin ^{2}\nu _{1}t+\sin ^{2}\nu _{2}t
}{2\sinh t}+\frac{\cos 2\nu _{1}t\cos 2\nu _{2}t-1}{2\sinh 2t}\right] dt- 
\nonumber \\
&&\qquad \qquad \qquad \qquad \qquad \qquad \qquad \frac{\gamma +2\log
2+\log M\varepsilon }{4}\ ,  \label{2.11}
\end{eqnarray}
\begin{eqnarray}
&&\frac{\cosh \pi \nu _{1}\cosh \pi \nu _{2}}{\pi ^{2}}\int\limits_{M
\varepsilon }^{\infty }K_{\frac{1}{2}+i\nu _{1}}^{2}\left( x\right) K_{\frac{
1}{2}+i\nu _{2}}^{2}\left( x\right) xdx=  \nonumber \\
&&\int\limits_{0}^{\infty }\left[ \frac{\sinh \left( 1+2i\nu _{1}\right)
t\sinh \left( 1+2i\nu _{2}\right) t}{2\sinh 2t}-\frac{1}{4}e^{2i\left( \nu
_{1}+\nu _{2}\right) t}\right] dt+  \nonumber \\
&&\qquad \qquad \frac{\gamma \left( \frac{1}{2}+i\nu _{1}\right) \gamma
\left( \frac{1}{2}+i\nu _{2}\right)\left( \frac{2}{M\varepsilon }\right) 
^{2i\left( \nu _{1}+\nu _{2}\right) } -1}{8i\left( \nu _{1}+\nu _{2}\right) } \ ,
\label{2.12}
\end{eqnarray}
where $\gamma =0.577216\cdots $ is the Euler constant and 
\begin{equation}
\gamma \left( x\right) \equiv \frac{\Gamma \left( x\right) }{\Gamma \left(
1-x\right) }\ .  \label{2.25}
\end{equation}
Let us imagine that $x$ integration in (\ref{2.9}) with the help of (\ref
{2.11}) and (\ref{2.12}) is already performed. Then the resulting expression
can be represented as a sum of two parts 
\begin{equation}
I_{\varepsilon }^{1}\left( a\right) =\sum\limits_{\nu _{1},\nu _{2}\in {\cal 
N}_{\varepsilon }^{+}}\frac{1}{\rho \left( \nu _{1}\right) \rho \left( \nu
_{2}\right) }\left[ {\LARGE i}\right] +\sum\limits_{\nu _{1},\nu _{2}\in 
{\cal N}_{\varepsilon }^{+}}\frac{1}{\rho \left( \nu _{1}\right) \rho \left(
\nu _{2}\right) }\left[ {\Large ii}\right] ,  \label{2.13}
\end{equation}
where in the second part, symbolically denoted as $\left[ ii\right] $, are
collected all the terms induced by the term 
\begin{equation}
\frac{\gamma \left( \frac{1}{2}+i\nu _{1}\right) \gamma \left( \frac{1}{2}
+i\nu _{2}\right)\left( \frac{2}{M\varepsilon }\right) ^{2i\left( \nu _{1}+
\nu _{2}\right) }  -1}{8i\left( \nu _{1}+\nu _{2}\right) }  \label{2.23}
\end{equation}
of the equation (\ref{2.12}). As the $\nu _{1}$, $\nu _{2}$ dependence of
terms collected in the remaining part $\left[ i\right] $ (in contrary to those
of $\left[ ii\right] $ ) are free of rapid , comparable with $\log
1/M\varepsilon $ frequency oscillations, it is safe to make replacement 
\begin{equation}
\sum\limits_{\nu _{1},\nu _{2}\in {\cal N}_{\varepsilon }^{+}}\frac{1}{\rho
\left( \nu _{1}\right) \rho \left( \nu _{2}\right) }\left[ {\Large i}\right]
\rightarrow \int\limits_{0}^{\infty }\int\limits_{0}^{\infty }d\nu _{1}d\nu
_{2}\left[ {\Large i}\right] \ .  \label{2.14}
\end{equation}
Surprisingly enough it is possible to factorize these integrals in such a
way, that they can be performed explicitly using the formula (for proof see
Appendix A) 
\begin{equation}
Im\int\limits_{0}^{\infty }d\nu \frac{e^{-2\pi \nu +i\pi \alpha +2i\nu t}}{
1+e^{-2\pi \nu +i\pi \alpha }}=\frac{1}{4t}-\frac{e^{-\alpha t}}{4\sinh t}\ .
\label{2.26}
\end{equation}
The sum over $\nu _{1}$, $\nu _{2}$ in second term of (\ref{2.13}) also can
be converted into the integrals provided one notices that (\ref{2.23}) is
exactly zero when $\nu _{1},\nu _{2}\in {\cal N}_{\varepsilon }$ , $\nu
_{1}+\nu _{2}\neq 0$ and is equal to $\pi \rho \left( \nu _{1}\right) /4$
(up to nonessential, finite in the $\varepsilon \rightarrow 0$ limit term)
if $\nu _{1}+\nu _{2}=0$, so that (\ref{2.23}) can be effectively replaced
by $\pi \delta \left( \nu _{1}+\nu _{2}\right) /4$. This leaves us with
elementary one dimensional integrals. As a result of above described
calculation for (\ref{2.9}) one obtains

\begin{eqnarray}
&&2\pi I_{\varepsilon }^{1}\left( a\right) =\pi \alpha \cot \frac{\pi \alpha 
}{2}-2+\frac{\alpha ^{2}}{2}\left( \psi \left( \frac{1}{2}\right) -\log
2\right) +  \nonumber \\
&&\int\limits_{0}^{\infty }dt\left[ \frac{\sinh ^{2}\alpha t}{\sinh 2t\sinh
^{2}t}+\frac{\alpha ^{2}\left( 2\cosh t-1\right) }{\sinh 2t}-\frac{\alpha
\sinh \alpha t}{2\sinh ^{2}t}+\right.  \label{2.15} \\
&&\left. \frac{2\cosh t\sinh ^{4}\frac{\alpha }{2}t}{\sinh ^{3}t}+\frac{
\cosh \alpha t-1}{\sinh ^{2}t}-\frac{\sinh ^{2}\alpha t}{\sinh 2t}-
\frac{\alpha ^{2} e^{-2t}}{2t}\right] \ ,
\nonumber
\end{eqnarray}
where $\psi \left( x\right) $ is the logarithmic derivative of the $\Gamma $
-function. In (\ref{2.15}) and later on we set 
\begin{equation}
\alpha =2\sqrt{2}a\ .  \label{2.24}
\end{equation}
Using the table of integrals presented at the end of Appendix A it is not
difficult to perform $t$ integration too 
\begin{eqnarray}
&&2\pi I_{\varepsilon }^{1}\left( a\right) =-\frac{1}{2}\psi \left( \frac{1}{
2}\right) -1+\frac{\pi \alpha }{2}\cot \frac{\pi \alpha }{2}+\alpha
^{2}\left( \frac{1}{4}-\log 2\right) +  \nonumber \\
&&\frac{\alpha ^{2}}{4}\left( \psi \left( \frac{\alpha }{2}\right) +\psi
\left( -\frac{\alpha }{2}\right) \right) +\frac{1-\alpha ^{2}}{4}\left( \psi
\left( \frac{1+\alpha }{2}\right) +\psi \left( \frac{1-\alpha }{2}\right)
\right) .  \label{2.16}
\end{eqnarray}
Now let us turn to the computation of the renormalization factor $Z$ in ( 
\ref{2.5}). If $\varepsilon $ is small enough we can split the region $
\left| z\right| \geq \varepsilon $ into two pieces by the circle $\left|
z\right| =\widetilde{\varepsilon }$ with some $\widetilde{\varepsilon }$
satisfying the conditions 
\begin{equation}
\log \frac{\widetilde{\varepsilon }}{\varepsilon }\gg 1;\quad \log M
\widetilde{\varepsilon }\ll 1\,  \label{2.17}
\end{equation}
so that inside the first region ${\it U}_{1}=\left\{ z;\varepsilon \leq
\left| z\right| \leq \widetilde{\varepsilon }\right\} $ the theory is nearly
conformal invariant and at the same time in the region ${\it U}_{2}=\left\{
z;\left| z\right| >\widetilde{\varepsilon }\right\} $ the influence of the
boundary at $\left| z\right| =\varepsilon $ could be neglected. Note that in
contrary to the region ${\it U}_{2}$ where the regularization prescription ( 
\ref{1.21}) is standard and in Cartesian coordinates transforms to the usual
normal ordering , due to the influence of the boundary in region ${\it U}
_{1} $ the interaction term of the Hamiltonian (\ref{1.20}) results in an
extra multiplicative renormalization of the field $\exp ia\varphi $ (besides
the usual charge renormalization $a_{r}=\left( 1-g/2\pi \right) a$ ). The
actual computation of the renormalization constant $Z$ is significantly
simplified owing to the existence of conformal invariance inside the region $
{\it U}_{1} $. As usual in CFT it is convenient to use radial quantization 
\cite{BPZ}. let us denote by $\left| B,a\right\rangle $ the boundary state 
\cite{Cardy} corresponding to the boundary conditions (\ref{1.9}) and
belonging to the conformal family \cite{BPZ} of the state $\left|
a\right\rangle =$ $\exp ia\varphi \left( 0\right) \left| 0\right\rangle $.
Note that during the evaluation from $\varepsilon $ to $\widetilde{
\varepsilon }$, the state $\varepsilon ^{-2a^{2}}\left| B,a\right\rangle $
approaches to $\widetilde{\varepsilon }^{-2a^{2}}\left| a\right\rangle $
thus correctly imitating the insertion of the field $\exp ia\varphi \left(
0\right) $. This consideration makes transparent the appearance of the
factor $\varepsilon ^{-2a^{2}}$ in (\ref{2.5}). It is not difficult to see
that the renormalization factor $Z$ is given by 
\begin{equation}
Z-1=-4\pi g\int\limits_{\varepsilon }^{\widetilde{\varepsilon }}\left\langle
a\right| N\left( \psi _{L}^{\dagger }\psi _{L}\psi _{R}^{\dagger }\psi
_{R}\right) \varepsilon ^{L_{0}+\overline{L}_{0}}\left| B,a\right\rangle
\left| z\right| d\left| z\right| -\frac{g\alpha ^{2}}{4\pi }\log \frac{
\widetilde{\varepsilon }}{\varepsilon }\ .  \label{2.18}
\end{equation}
Here and in what follows we take the Fermi fields in initial coordinates $z$
, $\overline{z}$ (i.e. the transformation (\ref{1.5}) is not applied). The
operator $\varepsilon ^{L_{0}+\overline{L}_{0}}$ ($L_{0}$, $\overline{L}_{0}$
are the Virasoro generators) is included in (\ref{2.18}) to take into
account that the boundary state $\left| B,a\right\rangle $ is associated to
the circle $\left| z\right| =\varepsilon $. The second term in (\ref{2.18})
subtracts the contribution of the charge renormalization. In Appendix B we
have presented the details of the computation of the matrix element included
in (\ref{2.18}). Inserting (\ref{B6}) into (\ref{2.18}) and performing
integration with the help of (\ref{A22}) we obtain 
\begin{equation}
Z-1=-\frac{g}{2\pi }\left( 1-\frac{\pi \alpha }{2}\cot \frac{\pi \alpha }{2}
\right) .  \label{2.19}
\end{equation}
As it should be expected the choice of $\widetilde{\varepsilon }$ satisfying
the conditions (\ref{2.17}) has no effect on the value of $Z$. Now taking
into account (\ref{2.5}), (\ref{2.22}), (\ref{2.16}) and (\ref{2.19}) we can
write down a final expression for the expansion of VEV up to linear over $g$
terms 
\begin{eqnarray}
&&\frac{\left\langle e^{ia\varphi \left( 0\right) }\right\rangle }{\left\langle
e^{ia\varphi \left( 0\right) }\right\rangle \mid _{g=0}}=1+\frac{g}{8\pi }
\left[ -2\psi \left( \frac{1}{2}\right) +\alpha ^{2}\left( 1-4\log 2\right)
+\right.  \nonumber \\
&&\left. \alpha ^{2}\left( \psi \left( \frac{\alpha }{2}\right) +\psi \left(
-\frac{\alpha }{2}\right) \right) +\left( 1-\alpha ^{2}\right) \left( \psi
\left( \frac{1+\alpha }{2}\right) +\psi \left( \frac{1-\alpha }{2}\right)
\right) \right]  \nonumber \\
&&\qquad \qquad \qquad \qquad \qquad \qquad \qquad \qquad \qquad \qquad
\qquad \qquad \quad +O\left( g^{2}\right) .  \label{2.21}
\end{eqnarray}
It is not difficult to check that (\ref{2.21}) exactly coincides with the
expression which one obtains directly expanding Lukyanov-Zamolodchikov
formula \cite{LukZam}.

\vspace{0.5cm} 
\appendix{\large {\bf Acknowledgements}}
\vspace{0.25cm}

The author is grateful to his colleges H.Babujian and A.Sedrakyan for 
interesting discussions. This research is partially supported by INTAS grant 96-524.

\vspace{0.5cm} \setcounter{equation}{0} \appendix{\large Appendix A} 
\renewcommand{\theequation}{A.\arabic{equation}} \vspace{0.25cm}

It appears that the Hankel-transforms are appropriate tools allowing us to
perform the integration over $x$ in (\ref{2.9}). In polar coordinates the
Hankel-transforms play the same role, as the ordinary Fourier-transforms in
the Cartesian one.

Let me briefly recall the main formulae concerning to the Hankel-transforms
(for details see \cite{BatErd} and references therein). The $\nu $-th order
( $\nu >-1$ ) direct and inverse Hankel-transforms of the function $f(x)$
defined on $(0,\infty )$ are given by 
\begin{equation}
f\left( x\right) =\int\limits_{0}^{\infty }J_{\nu }\left( sx\right) 
\widetilde{f_{\nu }}\left( s\right) sds,  \label{A1}
\end{equation}
\begin{equation}
\widetilde{f_{\nu }}\left( s\right) =\int\limits_{0}^{\infty }J_{\nu }\left(
sx\right) f\left( x\right) xdx,  \label{A2}
\end{equation}
where $J_{\nu }$ is the Bessel function. In complete analogy with the case
of Fourier-transform, it follows from (\ref{A1}), (\ref{A2}), that the
''scalar product'' of any two functions\footnote{
Since the functions we are dealing with are regular in the interval $
(0,\infty )$, the only thing one has to care of is the convergence of
integrals at the extreme points $0$ and $\infty $.} $f(x)$, $g(x)$ in $x$
representation coincides with the ''scalar product'' of their images in $s$
representation: 
\begin{equation}
\int\limits_{0}^{\infty }f\left( x\right) g\left( x\right)
xdx=\int\limits_{0}^{\infty }\widetilde{f_{\nu }}\left( s\right) \widetilde{
g_{\nu }}\left( s\right) sds.  \label{A3}
\end{equation}
The starting point of our consideration is the formula \cite{BatErd} 
\begin{eqnarray}
\left[ K_{\nu }\left( x\right) \right] ^{2}\sin \left( \nu \pi \right) 
&=&\pi \int\limits_{0}^{\infty }J_{0}\left( 2x\sinh t\right) \sinh \left(
2\nu t\right) dt,  \label{A4} \\
\left|Re\nu \right|  &<&\frac{3}{4}\ ;\ x>0,  \nonumber
\end{eqnarray}
which obviously can be considered (after substitution $s=\sinh t$) as the $0$
-th order Hankel-transform of the function $\left[ K_{\nu }\left( x\right)
\right] ^{2}$. Assuming $0<Re\nu _{1}<1/2$, $0<Re\nu _{2}<1/2$ and using (
\ref{A3}) one obtains 
\begin{eqnarray}
&&\frac{\sin \left( \nu _{1}\pi \right) \sin \left( \nu _{2}\pi \right) }{
\pi ^{2}}\int\limits_{0}^{\infty }\left[ K_{\nu _{1}}\left( x\right) \right]
^{2}\left[ K_{\nu _{2}}\left( x\right) \right]
^{2}xdx=\int\limits_{0}^{\infty }\frac{\sinh \left( 2\nu _{1}t\right) \sinh
\left( 2\nu _{2}t\right) }{2\sinh 2t}dt=  \nonumber \\
&&\qquad \qquad \qquad \qquad \qquad \qquad \frac{1}{16}\left[ \psi \left( 
\frac{1+\nu _{1}-\nu _{2}}{2}\right) +\psi \left( \frac{1-\nu _{1}+\nu _{2}}{
2}\right) \right. -  \nonumber \\
&&\qquad \qquad \qquad \qquad \qquad \qquad \qquad \left. \psi \left( \frac{
1+\nu _{1}+\nu _{2}}{2}\right) +\psi \left( \frac{1-\nu _{1}-\nu _{2}}{2}
\right) \right] ,  \label{A5}
\end{eqnarray}
where the second equality follows from (\ref{A22}). For our purposes we have
to restrict the range of integration in (\ref{A5}) to $[M\varepsilon ,\infty
)$ and substitute $\nu _{1}\rightarrow 1/2+i\nu _{1}$, $\nu _{1}\rightarrow
1/2+i\nu _{2}$. This has to be done carefully, since for such values of
parameters the integral (\ref{A5}) doesn't converge at $x\rightarrow 0$. To
avoid this difficulty, one first has to subtract the integral 
\begin{eqnarray}
&&\frac{\sin \left( \nu _{1}\pi \right) \sin \left( \nu _{2}\pi \right) }{
\pi ^{2}}\int\limits_{0}^{M\varepsilon }\left[ K_{\nu _{1}}\left( x\right)
\right] ^{2}\left[ K_{\nu _{2}}\left( x\right) \right] ^{2}xdx=  \nonumber \\
&&\qquad \frac{1}{2\left( 1-\nu _{1}-\nu _{2}\right) }\ \left( \frac{2}{
M\varepsilon }\right) ^{2\left( \nu _{1}+\nu _{2}-1\right) }\left( 1+O\left(
\varepsilon ^{2}\right) \right)   \label{A6}
\end{eqnarray}
from (\ref{A5}) and only after it to make above mentioned substitutions $\nu
_{1}\rightarrow 1/2+i\nu _{1}$, $\nu _{2}\rightarrow 1/2+i\nu _{2}$ . Note
that in (\ref{A6}) we have used the following expansion of $K_{\nu }^{2}\left(
x\right) $ near $0$: 
\begin{equation}
K_{\nu }^{2}\left( x\right) =\frac{\pi \gamma \left( \nu \right) }{4\sin
\left( \nu \pi \right) }\left( \frac{2}{x}\right) ^{2\nu }\left( 1+O\left(
x^{2}\right) \right) .  \label{A7}
\end{equation}
As a result one easily arrives at (\ref{2.12}).

The proof of (\ref{2.11}) we begin with the formula 
\begin{equation}
\frac{\sin \left( \nu \pi \right) }{\pi }K_{\nu }\left( x\right) K_{1-\nu
}\left( x\right) =\frac{1}{2x}-\int\limits_{0}^{\infty }\left[ J_{1}\left(
2x\sinh t\right) \cosh \left( \left( 2\nu -1\right) t\right) \right] dt,
\label{A8}
\end{equation}
which can be verified differentiating (\ref{A4}) over $x$ and using the
identities 
\begin{eqnarray}
K_{\nu }^{^{\prime }}\left( x\right) &=&-K_{1-\nu }\left( x\right) -\frac{
\nu }{x}K_{\nu }\left( x\right) , \nonumber \\
J_{0}^{^{\prime }}\left( x\right) &=&-J_{1}\left( x\right) \label{A29}.
\end{eqnarray}
Since the image of the function $1/x$ under the first order Hankel-transform
is equal to $1$, the equation (\ref{A8}) (after substitution $\nu
\rightarrow 1/2+i\nu $) can be rewritten as 
\begin{equation}
\frac{\cosh \left( \nu \pi \right) }{\pi }\left| K_{_{\frac{1}{2}+i\nu
_{1}}}\left( x\right) \right| ^{2}=\int\limits_{0}^{\infty }J_{1}\left(
2x\sinh t\right) \left( \cos 2\nu t-\cosh t\right) dt.  \label{A9}
\end{equation}
Unfortunately (\ref{A9}) can not be used directly to calculate integral (\ref
{2.11}) because of the singularity at $x=0$: 
\begin{equation}
\frac{\cosh \left( \nu \pi \right) }{\pi }\left| K_{\nu }\left( x\right)
\right| ^{2}=\frac{1}{2x}\left( 1+O\left( x^{2}\right) \right) .  \label{A10}
\end{equation}
To overcome this difficulty we replace the functions 
\begin{equation}
\frac{\cosh \left( \nu \pi \right) }{\pi }\left| K_{_{\frac{1}{2}+i\nu
_{1}}}\left( x\right) \right| ^{2},\ \frac{\cosh \left( \nu \pi \right) }{
\pi }\left| K_{_{\frac{1}{2}+i\nu _{2}}}\left( x\right) \right| ^{2}
\label{A11}
\end{equation}
with their regularized versions 
\begin{equation}
\frac{\cosh \left( \nu \pi \right) }{\pi }\left| K_{_{\frac{1}{2}+i\nu
_{1}}}\left( x\right) \right| ^{2}-\frac{e^{-\Lambda x}}{2x}\ ,\ \frac{\cosh
\left( \nu \pi \right) }{\pi }\left| K_{_{\frac{1}{2}+i\nu _{2}}}\left(
x\right) \right| ^{2}-\frac{e^{-\Lambda x}}{2x}\ ,  \label{A12}
\end{equation}
where $\Lambda $ is some large positive cut-off. The first order Hankel
images of the functions (\ref{A12}) can be obtained using the formulae 
\begin{equation}
\frac{e^{-\Lambda x}}{2x}=\frac{1}{2}\int\limits_{0}^{\infty }J_{1}\left(
sx\right) \left( 1-\frac{\Lambda }{\sqrt{\Lambda ^{2}+s^{2}}}
\right) ds  \label{A13}
\end{equation}
and (\ref{A9}). Then one first applies (\ref{A3}) to calculate the
regularized version of the integral (\ref{2.11}) over the full interval $
(0,\infty )$ , next subtracts the integral over the interval $
(0,M\varepsilon )$ (as $M\varepsilon \ll 1$, one simply applies the
asymptotic formula (\ref{A10}) ) and, afterwards gets rid of the cut-off $
\Lambda $ tending it to $\infty $. The final result is the equation (\ref
{2.11}). Note that using (\ref{A22}), integration over $t$ on the right hand
side of (\ref{2.11}) may be expressed via the $\psi $ function. For the sake
of completeness let us present here this expression too 
\begin{eqnarray}
&&\frac{\cosh \pi \nu _{1}\cosh \pi \nu _{2}}{\pi ^{2}}\int\limits_{
\varepsilon }^{\infty }\left| K_{\frac{1}{2}+i\nu _{1}}\left( x\right)
\right| ^{2}\left| K_{\frac{1}{2}+i\nu _{2}}\left( x\right) \right| ^{2}xdx=
\nonumber \\
&&\qquad \qquad \sum\limits_{\sigma =\pm 1}\left[ \frac{1}{8}\left( \psi
\left( \frac{1}{2}+i\sigma \nu _{1}\right) +\psi \left( \frac{1}{2}+i\sigma
\nu _{2}\right) \right) -\right.  \nonumber \\
&&\qquad \left. \frac{1}{16}\left( \psi \left( \frac{1}{2}+i\sigma \frac{\nu
_{1}+\nu _{2}}{2}\right) +\psi \left( \frac{1}{2}+i\sigma \frac{\nu _{1}-\nu
_{2}}{2}\right) \right) \right] - \nonumber \\
&&\qquad \qquad \qquad \qquad \qquad \qquad \qquad \frac{1}{4}\ \left(
1+O\left( \varepsilon ^{2}\right) \right) \log \varepsilon \ .  \label{A27}
\end{eqnarray}

Now let us sketch the proof of the equation (\ref{2.26}). Consider the
function $L\left( a,b\right) $ ($a$, $b$ are real variables and $\left|
a\right| <\pi $) defined by the integral 
\begin{equation}
L\left( a,b\right) =\int\limits_{0}^{\infty }e^{ibx}\log \left(
1+e^{-x+ia}\right) dx  \label{A14}
\end{equation}
and the contour integral 
\begin{equation}
J\left( {\it C};b\right) =\int\limits_{{\it C}}e^{ibz}\log \left(
1+e^{-z}\right) dz  \label{A15}
\end{equation}
for any contour ${\it C}$ on the complex $z$-plain. Let ${\it C}_{1,a}$, $
{\it C}_{2,a}$ be the contours 
\begin{eqnarray}
&&{\it C}_{1,a}:\quad z=-i\pi +it;\quad 0<t\le \pi -a,  \nonumber \\
&&{\it C}_{2,a}:\quad z=-ia+t;\quad t>0\ .  \label{A28}
\end{eqnarray}
As the integration contour can be freely deformed inside analyticity domain
it is clear that 
\begin{equation}
J\left( {\it C}_{1,a};b\right) +J\left( {\it C}_{2,a};b\right) =J\left( {\it 
C}_{2,\pi };b\right) =J\left( {\it C}_{1,-\pi };b\right) +J\left( {\it C}
_{2,-\pi };b\right) .  \label{A16}
\end{equation}
Taking into account the evident relations 
\begin{eqnarray}
&&J\left( {\it C}_{2,a},b\right) =e^{ab}L\left( a,b\right) ,  \nonumber \\
&&L\left( \pi ,b\right) =L\left( -\pi ,b\right) \label{A17} 
\end{eqnarray}
we obtain 
\begin{equation}
J\left( {\it C}_{1,a};b\right) +e^{ab}L\left( a,b\right) =\frac{e^{\pi b}}{
e^{\pi b}-e^{-\pi b}}\ J\left( {\it C}_{1,-\pi };b\right) .  \label{A18}
\end{equation}
Thus 
\begin{eqnarray}
&&e^{ab}\left( L\left( a,b\right) +L\left( -a,-b\right) \right) =-J\left( 
{\it C}_{1,a};b\right) -J\left( {\it C}_{1,-a};-b\right) +  \nonumber \\
&&\frac{e^{\pi b}}{2\sinh \pi b}\ J\left( {\it C}_{1,-\pi };b\right) -\frac{
e^{-\pi b}}{2\sinh \pi b}\ J\left( {\it C}_{1,-\pi };-b\right) .  \label{A19}
\end{eqnarray}
But an easy exercise shows that the r.h.s. of (\ref{A19}) is equal to 
\begin{equation}
\frac{e^{-\pi b}}{2\sinh \pi b}\int\limits_{-\pi }^{\pi
}xe^{-bx}dx+\int\limits_{-a}^{\pi }xe^{-bx}dx=e^{ab}\left( \frac{1-ab}{b^{2}}
-\frac{\pi e^{-ab}}{b\sinh \pi b}\right) .  \label{A20}
\end{equation}
So we have obtained 
\begin{equation}
{\it Re}\int\limits_{0}^{\infty }e^{ibx}\log \left( 1+e^{-x+ia}\right) dx=
\frac{1-ab}{2b^{2}}-\frac{\pi e^{-ab}}{2b\sinh \pi b}\ .  \label{A21}
\end{equation}
Differentiating the equation (\ref{A21}) over $a$ and changing the notations
in an obvious way we explicitly arrive at (\ref{2.26}).

We'll complete this section presenting a table of integrals, which has been
used when performing $t$ integration in (\ref{2.15}): 
\begin{equation}
\int\limits_{0}^{\infty }\frac{\cosh \alpha t-1}{\sinh ^{2}t}dt=1-\frac{\pi
\alpha }{2}\cot \frac{\pi \alpha }{2}  \label{A22}
\end{equation}
\begin{equation}
\int\limits_{0}^{\infty }\frac{\cosh \alpha t-1}{\sinh t}dt=\psi \left( 
\frac{1}{2}\right) -\frac{1}{2}\left( \psi \left( \frac{1+\alpha }{2}\right)
+\psi \left( \frac{1-\alpha }{2}\right) \right) ,  \label{A23}
\end{equation}
\begin{eqnarray}
&&\int\limits_{0}^{\infty }\left( \frac{\sinh ^{2}\alpha t}{\sinh 2t\sinh
^{2}t}-\frac{\alpha ^{2}}{\sinh 2t}\right) dt=\frac{3\alpha ^{2}+\left(
2\alpha ^{2}-1\right) \psi \left( \frac{1}{2}\right) }{4}-  \nonumber \\
&&\frac{\alpha ^{2}}{8}\left( \psi \left( \frac{\alpha }{2}\right) +\psi
\left( -\frac{\alpha }{2}\right) \right) +\frac{1-\alpha ^{2}}{8}\left( \psi
\left( \frac{1+\alpha }{2}\right) +\psi \left( \frac{1-\alpha }{2}\right)
\right) ,  \label{A24}
\end{eqnarray}
\begin{equation}
\int\limits_{0}^{\infty }\left( \frac{\alpha \sinh \alpha t}{\sinh ^{2}t}- 
\frac{\alpha ^{2}}{\sinh t}\right) dt=\frac{\alpha ^{2}}{2}\left( 2+2\psi
\left( \frac{1}{2}\right) -\psi \left( \frac{\alpha }{2}\right) -\psi \left(
-\frac{\alpha }{2}\right) \right) ,  \label{A25}
\end{equation}
\begin{equation}
\int\limits_{0}^{\infty }\frac{\cosh t\sinh ^{4}\frac{\alpha }{2}t}{\sinh
^{3}t}dt=\frac{\alpha ^{2}}{8}\left( \psi \left( \frac{\alpha }{2}\right)
+\psi \left( -\frac{\alpha }{2}\right) -\psi \left( \alpha \right) -\psi
\left( -\alpha \right) \right) .  \label{A26}
\end{equation}
Note that (\ref{A23}) is a direct consequence of the standard integral
representation for the function $\psi $ and that (\ref{A24})-(\ref{A26})
after some algebraic manipulations and integrations by part may be reduced
to (\ref{A23}).

\vspace{0.5cm} \setcounter{equation}{0} \appendix{\large Appendix B} 
\renewcommand{\theequation}{B.\arabic{equation}} \vspace{0.25cm} 

The usual mode decomposition of massless Fermi fields obeying the monodromy
relation (\ref{2.1}) takes the form 
\begin{eqnarray}
&&\psi _{L}\left( z\right) =\frac{1}{\sqrt{2\pi }}\sum\limits_{n\in {\cal Z}
- \frac{\alpha }{2}}\frac{c_{n}}{z^{n+1/2}};\ \psi _{L}^{\dagger }\left(
z\right) =\frac{1}{\sqrt{2\pi }}\sum\limits_{n\in {\cal Z}+\frac{\alpha }{2}
} \frac{c_{n}^{\dagger }}{z^{n+1/2}}\ ,  \nonumber \\
&&\psi _{R}\left( \overline{z}\right) =\frac{1}{\sqrt{2\pi }}
\sum\limits_{n\in {\cal Z}+\frac{\alpha }{2}}\frac{\widetilde{c}_{n}}{
\overline{z}^{n+1/2}};\ \psi _{R}^{\dagger }\left( \overline{z}\right) = 
\frac{1}{\sqrt{2\pi }}\sum\limits_{n\in {\cal Z}-\frac{\alpha }{2}}\frac{
\widetilde{c}_{n}^{\dagger }}{\overline{z}^{n+1/2}}\ ,  \label{B1}
\end{eqnarray}
where the operators $c_{n}$, $c_{n}^{\dagger }$, $\widetilde{c}_{n}$, $
\widetilde{c}_{n}^{\dagger }$ satisfy the anti commutation relations 
\begin{equation}
\left\{ c_{n},c_{m}^{\dagger }\right\} =\delta _{n+m,0},\quad \left\{ 
\widetilde{c}_{n},\widetilde{c}_{m}^{\dagger }\right\} =\delta _{n+m,0},
\label{B2}
\end{equation}
with other anti commutators being $0$. The state $\left| a\right\rangle $ is
defined by the condition that it is annihilated by the positive mode
operators and the defining property of the boundary state $\left|
B,a\right\rangle $ is 
\begin{equation}
\left( c_{n}-\widetilde{c}_{-n}\right) \left| B,a\right\rangle =\left(
c_{-n}^{\dagger }-\widetilde{c}_{n}^{\dagger }\right) \left|
B,a\right\rangle =0  \label{B3}
\end{equation}
for every $n\in {\cal Z}-\frac{\alpha }{2}$. Using (\ref{B1})-(\ref{B3}) it
is easy to find all nonzero two point functions 
\begin{eqnarray}
&&\left\langle a\right| \psi _{L}^{\dagger }\left( z_{1}\right) \psi
_{L}\left( z_{2}\right) \left| B,a\right\rangle =\frac{1}{2\pi \left(
z_{1}-z_{2}\right) }\left( \frac{z_{2}}{z_{1}}\right) ^{\alpha /2}, 
\nonumber \\
&&\left\langle a\right| \psi _{R}^{\dagger }\left( z_{1}\right) \psi
_{R}\left( z_{2}\right) \left| B,a\right\rangle =\frac{1}{2\pi \left( 
\overline{z}_{1}-\overline{z}_{2}\right) }\left( \frac{\overline{z}_{2}}{
\overline{z}_{1}}\right) ^{-\alpha /2},  \nonumber \\
&&\left\langle a\right| \psi _{L}^{\dagger }\left( z_{1}\right) \psi
_{R}\left( z_{2}\right) \left| B,a\right\rangle =\frac{1}{2\pi \left( z_{1} 
\overline{z}_{2}-1\right) }\left( z_{1}\overline{z}_{2}\right) ^{-\alpha /2},
\nonumber \\
&&\left\langle a\right| \psi _{R}^{\dagger }\left( z_{1}\right) \psi
_{L}\left( z_{2}\right) \left| B,a\right\rangle =\frac{1}{2\pi \left( 
\overline{z}_{1}z_{2}-1\right) }\left( \overline{z}_{1}z_{2}\right) ^{\alpha
/2}  \label{B4}
\end{eqnarray}
Using the first definition of the perturbing operator $N\left( \psi
_{L}^{\dagger }\psi _{L}\psi _{R}^{\dagger }\psi _{R}\right) $ (see the
first equality in (\ref{1.21})) and the ''propagators'' (\ref{B4}) it is
straightforward to find the matrix element 
\begin{equation}
\left\langle a\right| N\left( \psi _{L}^{\dagger }\psi _{L}\psi
_{R}^{\dagger }\psi _{R}\right) \left| B,a\right\rangle =\frac{1}{4\pi
^{2}\left| z\right| ^{2}}\left( \frac{\sinh ^{2}\left( \frac{\alpha }{2}\log
\left| z\right| \right) }{\sinh ^{2}\left( \log \left| z\right| \right) }- 
\frac{\alpha ^{2}}{4}\right) .  \label{B5}
\end{equation}
During all this consideration we have assumed the boundary state $\left|
B,a\right\rangle $ to be associated with the unite circle $\left| z\right|
=1 $, but simple scaling arguments allow us immediately to write down the
corresponding expression for the case, when $\left| B,a\right\rangle $ is
attached to the circle $\left| z\right| =\varepsilon $ as well: 
\begin{equation}
\left\langle a\right| N\left( \psi _{L}^{\dagger }\psi _{L}\psi
_{R}^{\dagger }\psi _{R}\right) \varepsilon ^{L_{0}+\overline{L}_{0}}\left|
B,a\right\rangle =\frac{1}{4\pi ^{2}\left| z\right| ^{2}}\left( \frac{\sinh
^{2}\left( \frac{\alpha \log \left| z\right| }{2\varepsilon }\right) }{\sinh
^{2}\left( \frac{\log \left| z\right| }{\varepsilon }\right) }-\frac{\alpha
^{2}}{4}\right) .  \label{B6}
\end{equation}

\end{document}